\documentclass[twocolumn,superscriptaddress,nobibnotes,footinbib,floatfix,aps,prl,citeautoscript,reprint]{revtex4-1}

\usepackage{graphicx}
\usepackage{epsfig}
\usepackage{subfigure}
\usepackage{color}
\usepackage{latexsym}
\usepackage{amsmath,bm,multirow}
\usepackage{amsfonts}
\usepackage{dcolumn}           
\usepackage{natbib}
\usepackage[colorlinks,linkcolor=blue,citecolor=green]{hyperref}
\usepackage{physics}

\newcommand{\etal}{\textit{et al.~}}

\usepackage[draft]{changes} 
\usepackage[normalem]{ulem}
\definechangesauthor[name={yi}, color=blue]{yi}

\newcommand{\nw}{Department of Materials Science and Engineering, Northwestern University, Evanston, IL 60208, USA}

\newcommand{\yale}{Department of Applied Physics, Yale University, New Haven, CT 06511, USA}
\newcommand{\esi}{Energy Sciences Institute, Yale University, West Haven, CT 06516, USA}

\newcommand{\CuSbS}{Cu$_{12}$Sb$_{4}$S$_{13}$}

\usepackage[draft]{changes} 
\usepackage[normalem]{ulem}
\definechangesauthor[name={Maria}, color=red]{Maria}
\definechangesauthor[name={Yi}, color=blue]{Yi}
\usepackage{soul}
\begin{document}
\title{Microscopic Mechanisms of Glass-Like Lattice Thermal Transport \\ in Cubic  Cu$_{12}$Sb$_{4}$S$_{13}$ Tetrahedrites}

\author{Yi Xia}
\email{yimaverickxia@gmail.com}
\affiliation{\nw}

\author{Vidvuds Ozoli\c{n}\v{s}}
\email{vidvuds.ozolins@yale.edu}
\affiliation{\yale}
\affiliation{\esi}

\author{Chris Wolverton}
\email{c-wolverton@northwestern.edu}
\affiliation{\nw}

\date{\today}

\begin{abstract}

Materials based on cubic tetrahedrites (\CuSbS) are useful thermoelectrics with unusual thermal and electrical transport properties, such as very low and nearly temperature-independent lattice thermal conductivity ($\kappa_{L}$). We explain the microscopic origin of the glass-like $\kappa_{L}$ in \CuSbS~by explicitly treating anharmonicity up to quartic terms for both phonon energies and phonon scattering rates. We show that the strongly unstable phonon modes associated with trigonally coordinated Cu atoms are anharmonically stabilized above approximately $100$~K and continue hardening with increasing temperature, in accord with experimental data. This temperature induced hardening effect reduces scattering of heat carrying acoustic modes by reducing the available phase space for three-phonon processes, thereby balancing the conventional $\propto T$ increase in scattering due to phonon population and yielding nearly temperature-independent $\kappa_{L}$. Furthermore, we find that very strong phonon broadening lead to a qualitative breakdown of the conventional phonon-gas model and modify the dominant heat transport mechanism from the particle-like phonon wave packet propagation to incoherent tunneling described by off-diagonal terms in the heat-flux operator, which are typically prevailing in glasses and disordered crystals. Our work paves the way to a deeper understanding of glass-like thermal conductivity in complex crystals with strong anharmonicity.

\end{abstract}

\maketitle


Materials exhibiting extreme thermal transport properties are of vital technological importance, with diverse  applications such as heat dissipation in electrical devices, thermal barrier coatings in gas turbines and jet engines, and thermal-to-electric energy conversion in thermoelectrics~\cite{Bell1457}. Improving the thermoelectric conversion efficiency by minimizing irreversible heat transport while preserving good electrical transport properties has been a central focus of research in thermoelectric materials~\cite{Snyder2008,Sootsman2009,Tritt2011}. This has inspired such breakthrough concepts as the paradigm of ``phonon glass/electron crystal (PGEC)"~\cite{PGEC,Nolas2001}, wherein ordered crystals can exhibit lattice thermal conductivity ($\kappa_{L}$) similar to that of amorphous solids or glasses, yet also possess good electronic properties~\cite{Beekman:2015aa}. Fundamental understanding of the microscopic mechanisms of thermal transport in such PGEC-type and related materials is nontrivial owing to the complex nature that bridges typical crystals and glasses~\cite{Einstein1911,Freeman1986,AllenPRL1989,Cahill1992,Allen1993disorder,shinde2006high,Voneshen:2013aa,Pailhes2014,tadano,Lory:2017aa,Tadano2018,Ikeda:2019aa,Minghui2019,Mukhopadhyay1455,Simoncelli2019,Isaeva2019,Luo:2020aa}; this persists to be a grand challenge in condensed matter and material physics~\cite{BeekmanReview}.

\begin{figure*}[htp]
	\includegraphics[width = 0.875\linewidth]{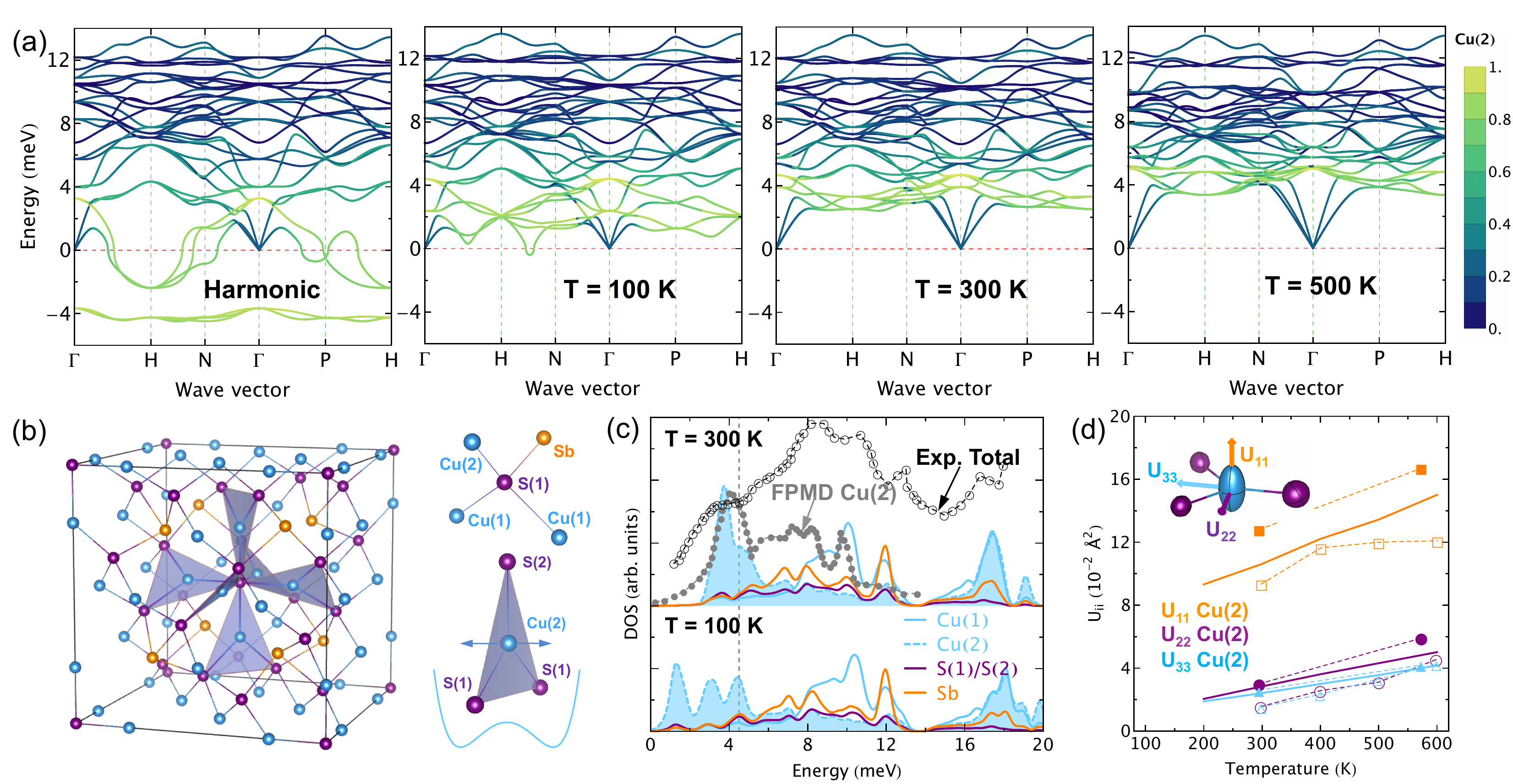}
	\caption{ 
	(a) Calculated phonon dispersions of \CuSbS~including anharmonic renormalization at finite temperatures ($T =  100$, $300$, and $500$~K) in comparison with those obtained from harmonic approximation. All bands are colored according to the amplitudes of the displacements of Cu(2) atoms. (b) Crystal structure of \CuSbS, the tetrahedron formed by Cu/Sb/S atoms and the trigonal coordination of Cu(2), wherein symmetrically inequivalent atoms are labeled as Cu(1), Cu(2), Sb, S(1) and S(2), respectively. A schematic double well potential energy surface is depicted for the out-of-plane displacements of Cu(2) atoms. (c) Calculated atom-decomposed phonon density of states (DOS) of \CuSbS~at $T = 100$~K (lower panel) and $T = 300$~K (upper panel), wherein the Cu(2) components are highlighted by filled colors, compared with the DOS of Cu(2) from first-principles molecular dynamics (FPMD) simulations (gray disks)~\cite{WeiLai2015} and the total DOS from inelastic neutron scattering measurements (black empty circles)~\cite{May2016}. (d) Calculated temperature-dependent anisotropic atomic displacement parameters ($U_{ii}$) of Cu(2) atoms compared with those obtained from FPMD simulations (empty shapes)~\cite{WeiLai2015} and experiments using Rietveld refinement (filled shapes)~\cite{Pfitzner1997}. The inset depicts the anisotropic thermal displacement ellipsoids of Cu(2) atoms.
	}
	\label{fig:StrPhonon}
\end{figure*}

Recently, \CuSbS~tetrahedrites have drawn vast attention because of their high thermoelectric performance and practical attractiveness due to easy synthesis and the fact that they are composed of nontoxic, earth-abundant elements~\cite{Suekuni_2012,Lu2013,SuekuniJAP,Heo:2014aa,Chetty2015PCCP,Lu:2015aa,Bouyrie2015}. They crystallize in a cubic sphalerite-like structure of $I\bar{4}3m$ symmetry at ambient conditions, with the Cu and S atoms occupying two symmetrically different crystallographic sites. The peculiarity of the structure is the coexistence of tetrahedral and trigonal planar coordinations for Cu(1) and Cu(2) atoms, respectively, as shown in Fig.~\ref{fig:StrPhonon}(b). Moreover, the displacements of Cu(2) atoms perpendicular to the trigonal plane have been reported to generate double-well potential energy surface (PES) [see Fig.~\ref{fig:StrPhonon}(b)] that leads to strong lattice instabilities and imaginary optical phonon modes at zero temperature [see Fig.~\ref{fig:StrPhonon}(a)]~\cite{Lu2013,csld}. These unstabe modes have been linked to the experimentally observed structural phase transition from cubic to tetragonal symmetry at approximately 85~K upon cooling~\cite{Suekuni_2012,Kitagawa2015,May2016}.

\begin{figure}[htp]
	\includegraphics[width = 1.0\linewidth]{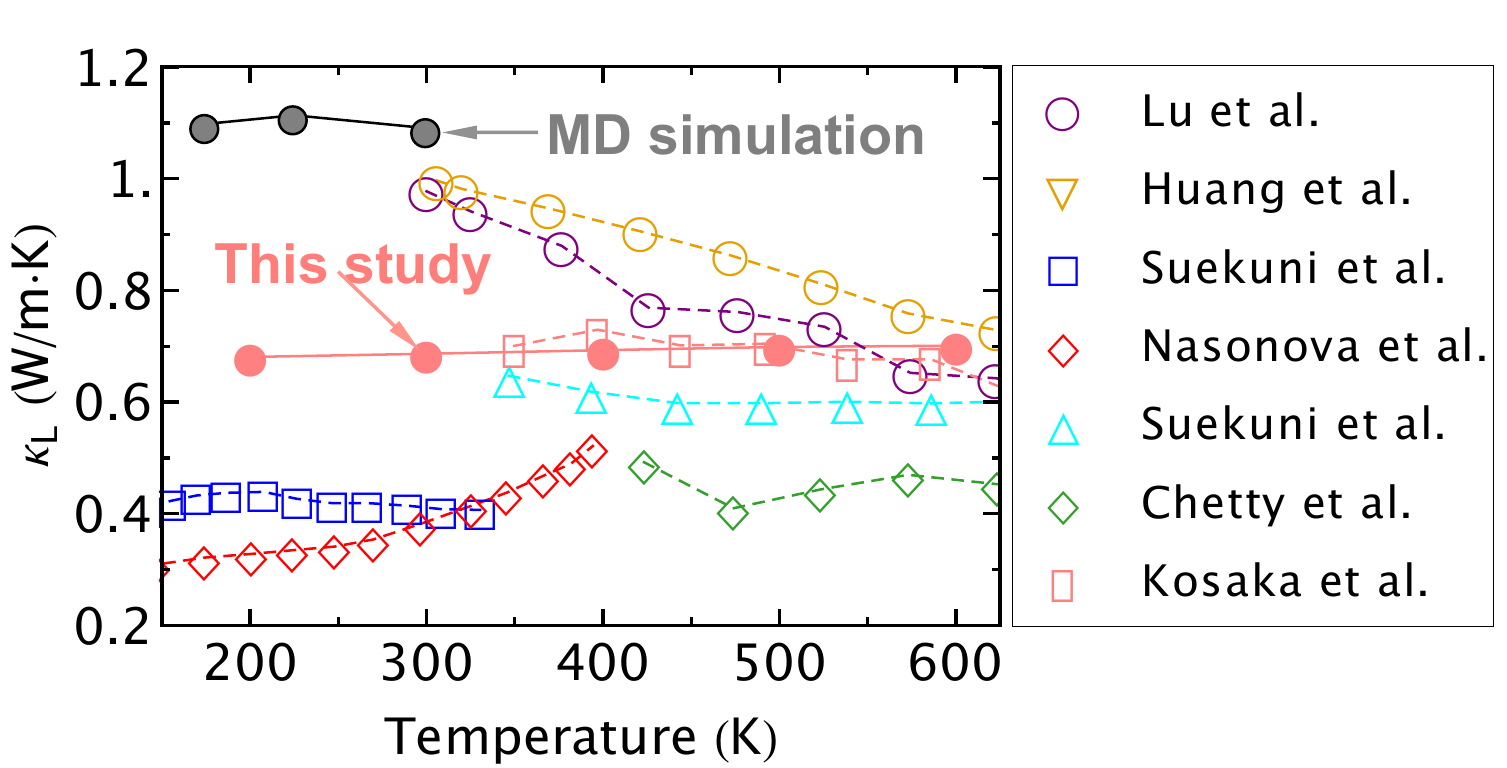}
	\caption{
	Calculated lattice thermal conductivity $\kappa_{3,4\text{ph}}^{P}+\kappa_{3,4\text{ph}}^{C}$ of \CuSbS~(red disks) in comparison with molecular dynamics (MD) simulations (black disks)~\cite{csld} and various experimental measurements on polycrystalline samples (empty shapes)~\cite{Lu2013,HUANG2018478,Suekuni_2012,Nasonova2016aa,Suekuni2018,Chetty2015PCCP,Kosaka2017PCCP}.
	}
	\label{fig:KappaAll}
\end{figure}

One of the most remarkable features of \CuSbS, an ordered high-symmetry crystal, is its lattice thermal conductivity $\kappa_{L}$ which mirrors amorphous compounds, displaying very low and nearly temperature-independent values of about 0.3 to 1.0 W/m$\cdot$K across a wide temperature range from 100~K to 600~K (see Fig.~\ref{fig:KappaAll})~\cite{Suekuni_2012,Lu2013,HUANG2018478,Nasonova2016aa,Suekuni2018,Chetty2015PCCP,Kosaka2017PCCP}. Numerous experimental and theoretical studies have attempted to explain the anomalous behavior of $\kappa_{L}$~\cite{lara2014,BouyriePCCP,WeiLai2015,May2016,Baoli2017}. They have suggested that electronic effects from lone $s^2$ pairs on the Sb$^{3+}$ ions combine with rattling-like vibrations of the Cu(2) ions to generate anharmonic forces, which play a key role in scattering heat-carrying acoustic phonons. However, a complete theoretical model that elucidates the microscopic heat transfer mechanisms underlying the glass-like $\kappa_{L}$ is still missing, mainly due to the challenges arising from the presence of massive lattice instabilities and severe high-order anharmonicity, which may invalidate the conventional phonon-gas model (PGM).


In this Letter, we investigate the lattice dynamics and thermal transport in cubic \CuSbS~by means of first-principles calculations based on the density functional theory (DFT). We construct a complete microscopic lattice dynamics model that is capable of rigorously accounting for both 3rd- and 4th-order anharmonicity and its effects on the phonon energies and phonon scattering rates. Using this model, we show that the unstable Cu(2) otpical modes are anharmonically stabilized at $T=110$~K and then continue hardening with increasing temperature. We develop an advanced theory of thermal transport in \CuSbS~that goes beyond the conventional PGM by taking into account both diagonal and off-diagonal terms of the heat-flux operator. We show that the almost temperature-independent $\kappa_{L}$ in \CuSbS~is a direct consequence of the strong anharmonic renormalization of the Cu(2) optical modes, which serves to decrease the available phase space for acoustic phonon scattering with increasing temperature. Moreover, we find that very strong phonon broadening lead to a surprising breakdown of PGM, which manifests in glass-like $\kappa_{L}$ dominated by the off-diagonal terms.


\begin{figure*}[htp]
	\includegraphics[width = 1.0\linewidth]{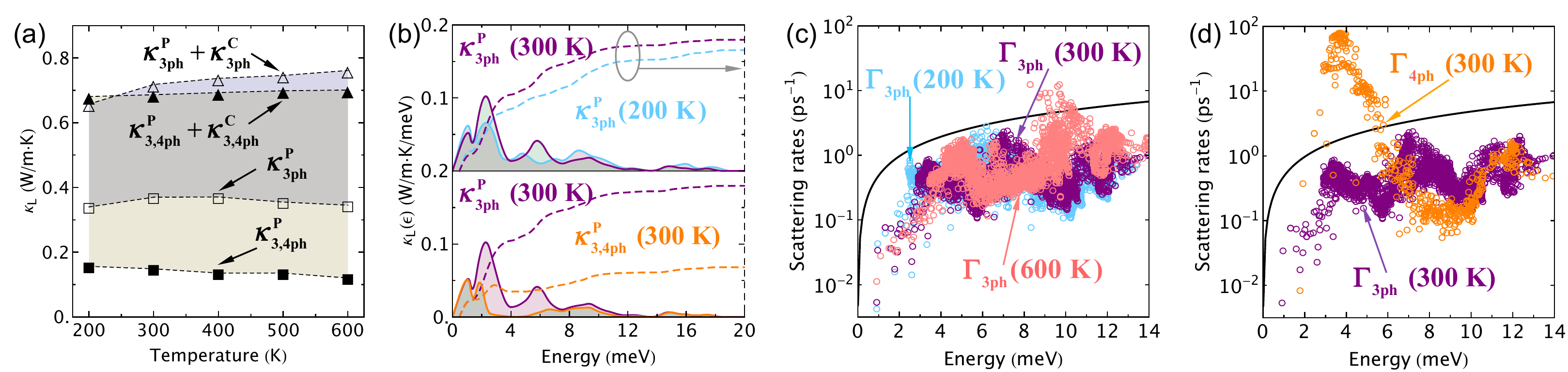}
	\caption{
	(a) Calculated temperature-dependent $\kappa_{L}$ considering only phonon populations' contribution ($\kappa_{L}^{P}$) and with additional coherences' contribution ($\kappa_{L}^{C}$). The empty shapes denote the values obtained considering only three-phonon (3ph) scattering, whereas the filled shapes further include four-phonon (4ph) scattering. (b) Comparisons of the cumulative (dashed lines)/differential (solid lines with filled colors) $\kappa_{L}^{P}$ obtained at $T =$200/300~K (upper panel) and with/without 4ph scattering at 300~K (lower panel). (c) Comparison of the 3ph scattering rates calculated at $T = 200$~K (blue dots), $T = 300$~K (purple dots) and $T = 600$~K (pink dots), respectively. (d) Comparison of the 3ph scattering rates (purple dots) and the 4ph scattering rates (orange circles) at $T = 300$~K. The solid black line in (c) and (d) indicate the scattering rate calculated by assuming that it equals twice the phonon frequency.
	}
	\label{fig:KappaAna}
\end{figure*}

\begin{figure}[htp]
	\includegraphics[width = 1.0\linewidth]{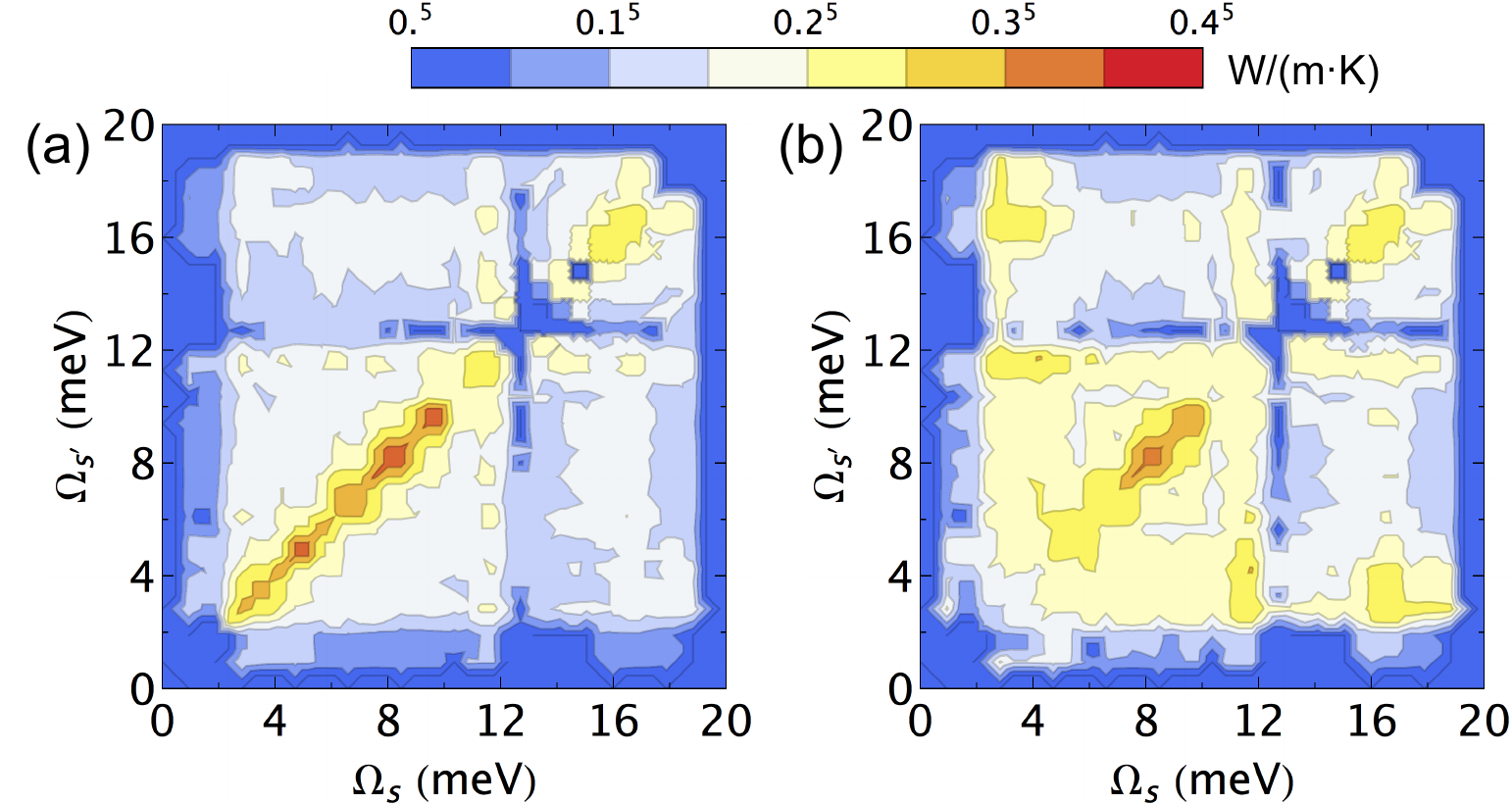}
	\caption{
	Contour plots of the coherences' part (off-diagonal terms) of the total lattice thermal conductivity associated with various pairs of phonon frequencies ($\Omega_{s}$ and $\Omega_{s^{\prime}}$) at $T = 300$~K. (a) is obtained considering only 3ph scatterings, whereas (b) includes additional 4ph scatterings.
	}
	\label{fig:KappaMap}
\end{figure}



We use the self-consistent phonon (SCPH) theory to accurately treat the phonon instabilities and anharmonicity arising from the Cu(2) double-well PES. In SCPH, the anharmonic phonon energies are obtained from the poles of the Green's function~\cite{choquard1967anharmonic,Cowley1968,Werthamer1970,mahan2000many,Errea2011,Tadano2015,XiaAgGeP2019}. Considering only the first-order contribution to the phonon self-energy from the quartic anharmonicity, the SCPH equation reads
\begin{equation}
\label{eq:SCPH}
\Omega^2_{q} = \omega^2_{q}+\frac{\hbar}{4} \sum_{q^{\prime}} \frac{V^{(4)}(q,-q,q^{\prime},-q^{\prime})}{\Omega_{q^{\prime}}} \left[ 1+2n\left(\Omega_{q^{\prime}}\right) \right],
\end{equation}
where $\omega$ is the harmonic phonon energy, $\Omega$ is the renormalized energy including anharmonic effects, and $q$ is a composite index of the phonon branch $s$ and wave vector $\mathbf{q}$. The quartic anharmonicity is represented using the Fourier-transformed 4th-order interatomic force constants (IFCs) $V^{(4)}$, while the temperature effects are reflected in the phonon population $n$, following the Bose-Einstein statistics. On top of the SCPH results, we further include phonon energy shifts from cubic anharmonicity in a perturbative manner~\cite{Maradudin1962,Cowley1968,Vanderbilt1991,Ribeiro2018}, which is found to substantially soften phonons in \CuSbS. We refer the readers to the Supplemental Material (SM)~\cite{SM} for more technical details and DFT calculations~\footnote{We performed DFT~\cite{dft} calculations using the Vienna {\it Ab\ Initio\/} Simulation Package (VASP)~\cite{Vasp1, Vasp2, Vasp3, Vasp4}, which employed the projector-augmented wave (PAW)~\cite{paw} method in conjunction with the Perdew-Burke-Ernzerhof version of the generalized gradient approximation (GGA)~\cite{gga} for the exchange-correlation functional~\cite{dft}. We extracted harmonic interatomic force constants (IFCs) using the finite-displacement approach implemented in Phonopy~\cite{Togo20151} and anharmonic IFCs up to quartic terms using compressive sensing lattice dynamics (CSLD)~\cite{csld,csldlong,csld2}, following the procedure detailed in Ref.[\onlinecite{csldlong}]. We have implemented and solved the SCPH equation using the ShengBTE package~\cite{shengbte,wuli2012} (see Supplemental Material~\cite{SM}, which includes additional Refs.[\onlinecite{TadanoJPSJ,Bianco2017,Lazzeri2003,TianliJAP,yixiagete2018}])}.

Upon including temperature effects, we find that the renormalized phonon energies are all stabilized above $T=$~200~K, as shown in Fig.~\ref{fig:StrPhonon}(a). The tiny unstable phonon energies present at $T=$~100~K can be removed by slightly increasing the temperature, making the cubic phase dynamically stable above $T=$~110~K (see Fig.S3 in SM~\cite{SM}). This value is qualitatively consistent with the experimental observation of a structural phase transition from tetragonal to cubic at $T=85$~K upon heating~\cite{Suekuni_2012}. By projecting the phonon spectrum and densities-of-states (DOS) onto different atomic sites [see Fig.~\ref{fig:StrPhonon}(a) and (c)], we find that the frequencies of the low-lying rattling-like vibrations associated with the Cu(2) atoms exhibit very large anharmonic renormalization and strong temperature dependence.  As shown in Fig.~\ref{fig:StrPhonon}c, the renormalized DOS at $T=$~300~K semi-quantitatively reproduces a vibrational Cu(2)-dominated peak near 4~meV that has been observed in both INS experiments~\cite{May2016} and first-principles molecular dynamics (FPMD) simulations~\cite{WeiLai2015}. Another peculiar feature of the Cu(2) vibrations is the large and anisotropic atomic displacement parameters (ADPs). Again, our calculated values based on the renormalized phonons, as shown in Fig.~\ref{fig:StrPhonon}(d), unambiguously confirm the unusually large ADPs of Cu(2) perpendicular to the trigonal plane, and lie in between experimental measurements~\cite{Pfitzner1997} and FPMD simulations~\cite{WeiLai2015}. We also find that phonon hardening leads to a weaker $T$ dependence of ADPs as compared to those obtained with fixed phonon dispersion at 200~K (see Fig.S4 in SM~\cite{SM}).

Having established a reliable lattice dynamics model, we proceed to explore the microscopic mechanisms of glass-like $\kappa_{L}$ in \CuSbS. Since the Pierels-Boltzmann theory~\cite{Peierls1929} (or PGM) only retains the diagonal terms of the harmonic heat-flux operator~\cite{Hardy1963,Semwal1972,Knauss1974,Srivastava1981}, whereas the off-diagonal contributions are no longer negligible in the presence of complex crystal, severe anharmonicity, or disorder~\cite{AllenPRL1989,Allen1993disorder,ShiLi2015,Lv:2016aa}, we employed a unified theory of thermal transport in crystals and glasses recently developed by Simoncelli, Marzari and Mauri~\cite{Simoncelli2019}. The resulting formula for $\kappa_{L}$ under single-mode approximation reads
\begin{equation}
\begin{split}
\kappa_{L} & = \frac{\hbar^2}{k_{B}T^2VN_{\mathbf{q}}} \sum_{\mathbf{q}}\sum_{s, s^{\prime}} \frac{ \Omega_{\mathbf{q}}^{s}+\Omega_{\mathbf{q}}^{s^{\prime}} }{2} \mathbf{v}_\mathbf{q}^{s,s^{\prime}} \otimes \mathbf{v}_\mathbf{q}^{s^{\prime},s}  \\
& \cdot \frac{ \Omega_\mathbf{q}^{s}n_\mathbf{q}^{s} ( n_\mathbf{q}^{s}+1) + \Omega_\mathbf{q}^{s^{\prime}}n_\mathbf{q}^{s^{\prime}} ( n_\mathbf{q}^{s^{\prime}}+1) }{ 4(\Omega_\mathbf{q}^{s^{\prime}}-\Omega_\mathbf{q}^{s})^2 + (\Gamma_\mathbf{q}^{s}+\Gamma_\mathbf{q}^{s^{\prime}})^2 }(\Gamma_\mathbf{q}^{s}+\Gamma_\mathbf{q}^{s^{\prime}}),
\end{split}
\label{eq:kappaall}
\end{equation}
where $\hbar$, $k_{B}$, $T$, $V$, $N_{\mathbf{q}}$ are respectively the reduced Plank constant, the Boltzmann constant, the absolute temperature, the volume of the unit cell and the number of sampled phonon wave vectors. And,   $\mathbf{v}_\mathbf{q}^{s,s^{\prime}}$ and $\Gamma_\mathbf{q}^{s}$ are respectively the generalized group velocity~\cite{Simoncelli2019} and scattering rate of a phonon mode indexed by wave vector $\mathbf{q}$ and branch $s$. When the second summation in the Eq.~(\ref{eq:kappaall}) is performed over the diagonal terms, namely $s=s^{\prime}$, Eq.~(\ref{eq:kappaall}) reduces to $\kappa^{P}_{L} = \frac{1}{VN_q} \sum_{q} C_q \mathbf{v}_q \otimes \mathbf{v}_q/\Gamma_{q}$ with $C_{q}$ the phonon mode heat capacity, which recovers the Pierels-Boltzmann theory in the conventional PGM and corresponds to the phonon populations' contribution to $\kappa_{L}$, denoted as $\kappa^{P}_{L}$.  The off-diagonal terms, which give the difference between $\kappa_{L}$ and $\kappa_{L}^{P}$, originate from the wave-like tunneling and loss of coherence between different vibrational eigenstates, thus being referred to as coherences' contribution $\kappa_{L}^{C}$ according to Ref.~\onlinecite{Simoncelli2019}. To estimate $\Gamma_{\mathbf{q}}^{s}$ on top of the renormalized phonons, we consider intrinsic three-phonon (3ph) and four-phonon (4ph) interactions~\cite{Maradudin1962,Debernardi1995,Tianli2016}, whose scattering rates are denoted as $\Gamma_{\text{3ph}}$ and $\Gamma_{\text{4ph}}$, respectively. We solved Eq.~(\ref{eq:kappaall}) with an in-house implementation of 4ph scattering rates on top of the renormalized phonons within the ShengBTE package~\cite{shengbte,wuli2012,yixiaprl2020} (see SM for technical details~\cite{SM}).

To reveal the dominant heat transfer mechanism, Fig.~\ref{fig:KappaAna}(a) shows the calculated $\kappa_{L}$ without and with coherences' contribution, namely $\kappa_{L}^{P}$ and $\kappa_{L}^{P}$+$\kappa_{L}^{C}$. Furthermore, to illustrate the effects of 4ph interactions, we also compute $\kappa_{L}$ using only 3ph scattering ($\kappa_{3\text{ph}}$) as well as including 4ph scattering ($\kappa_{3,4\text{ph}}$). The key findings inferred from Fig.~\ref{fig:KappaAna}(a) are as follows: (i) Considering only 3ph scattering, $\kappa^{P}_{3\text{ph}}$ displays very weak $T$-dependence while $\kappa_{3\text{ph}}^{C}$ becomes increasingly more important from 200~K to 600~K. (ii) Additional 4ph scattering leads to drastically reduced $\kappa_{3,4\text{ph}}^{P}$ but significantly enhanced $\kappa_{3,4\text{ph}}^{C}$, making the off-diagonal term dominant over the entire temperature range. (iii) The total $\kappa_{L} = \kappa^{P}_{3,4\text{ph}}+\kappa^{C}_{3,4\text{ph}}$ that accounts for both 3ph and 4ph scattering as well as non-Peierls-type off-diagonal contributions attains a $T$-independent value of  about 0.7 W/m$\cdot$K. Feature (i) is in striking contrast to the typical heat conduction in crystals, wherein $\kappa_{L}$ usually reaches a maximum at low $T$ ($<$~50~K) and then starts to decrease following a roughly $T^{-1}$ law at elevated $T$ ($>$~100~K) due to 3ph scattering~\cite{BeekmanReview}. 

By examining the key ingredients entering $\kappa_{L}$, namely, $C_{q}$, $\mathbf{v}_{q}$ and $\Gamma_{q}$, we find that the unusual $T$-dependence of $\kappa^{P}_{3\text{ph}}$ can be attributed to an anomalous behavior of the scattering rates $\Gamma_{3\text{ph}}$. As shown in Fig.~\ref{fig:KappaAna}(c), $\Gamma_{3\text{ph}}$ associated with some low-lying ($<$~4~meV) heat-carrying acoustic modes is surprisingly reduced when $T$ increases from 200~K to 600~K. Our analysis reveals that this anomalous behavior is caused by a balance between two effects: increasing $T$ leads to enhanced phonon population and thus larger $\Gamma_{3\text{ph}}$, while the hardening of the rattling-like Cu(2) vibrations result in reduced $\Gamma_{3\text{ph}}$ for the low-lying ($<$~4~meV) modes via decrease in their scattering phase space. This can readily explain the disappearance of large $\Gamma_{3\text{ph}}$ for modes with energies in the 2 to 4~meV interval when $T$ increases from 200~K to 600~K. The overall effects make $\kappa_{3\text{ph}}^{P}$ nearly independent of temperature. With additional 4ph scattering, $\kappa_{\text{3,4ph}}^{P}$ becomes significantly smaller than $\kappa_{\text{3ph}}^{P}$ [see Fig.~\ref{fig:KappaAna}(a)]  because of the giant values of $\Gamma_{\text{4ph}}$ [see Fig.~\ref{fig:KappaAna}(d)]. However, the nearly $T$-independence of $\kappa_{\text{3,4ph}}^{P}$ is found to largely persist even though 4ph scattering leads to a faster decay ($\propto T^{-2}$) of $\kappa_{L}$~\cite{Tianli2016}. This again shows the dramatic effect of the strong anharmonic frequency renormalization on the phonon scattering rates.

The large values of $\Gamma_{\text{4ph}}$ in conjunction with the dominant off-diagonal contributions to $\kappa_{L}$ indicate the breakdown of the traditional PGM description in \CuSbS~\footnote{Relatively weak phonon broadening is required to ensure the validity of PGM, while significant phonon broadening would induce coupling of vibrational modes arising from off-diagonal terms in the heat-flux operator, whose contribution to the thermal conductivity may dominate over the diagonal terms described by PGM.}. Specifically, considering only 3ph scattering, $\kappa_{L}^{C}$ comprises, percentage-wise, about 50\% of the total $\kappa_{L}$. With additional 4ph scattering effects, $\kappa_{L}^{C}$ significantly surpasses $\kappa_{L}^{P}$, comprising about 80\% of the total $\kappa_{L}$ over the entire temperature range. Such a dramatic change can be traced back to the mode-wise contributions to $\kappa_{L}^{C}$, as shown in Fig.~\ref{fig:KappaMap}. Considering only $\Gamma_{\text{3ph}}$, the largest contributions to $\kappa_{L}^{C}$ arise between nearby phonon modes with similar frequencies [see Fig.~\ref{fig:KappaMap}(a)]. In contrast, the giant $\Gamma_{\text{4ph}}$ induces coupling between phonon modes with large frequency differences, particularly for those involving the Cu(2) rattling modes. These results imply that heat transfer in \CuSbS~shares features with that in glasses, with the exception that in \CuSbS~it is due to strong anharmonicity, while in glasses the off-diagonal tunneling contributions arise from structural disorder. It is worth mentioning that scattering rates as high as those shown here, while rare, have also been reported in skutterudites that contain rattling atoms (YbFe$_{4}$Sb$_{12}$~\cite{Wuli2015}). We note that these very large scattering rates are likely overestimated due to the potential breakdown of perturbation theory. As such, our theoretical model might be further improved by going beyond perturbative evaluation of phonon scattering rates~\cite{Glensk2019,Aseginolaza2019}. Nevertheless, our results unambiguously demonstrate the important role of 4ph scattering in suppressing $\kappa_{L}^{P}$ while promoting $\kappa_{L}^{C}$.

We compare our results ($\kappa_{\text{3,4ph}}^{P}+\kappa_{\text{3,4ph}}^{C}$) with prior simulations and experiments in Fig.~\ref{fig:KappaAll}. The roughly $T$-independent character of $\kappa_{L}$ found in our study agrees with both earlier molecular dynamics (MD) simulations \footnote{The molecular dynamics (MD) is simulated using a force field based on a polynomial expansion of the DFT potential energy surface up to the sixth order. The thermal conductivity is computed using homogeneous nonequilibrium MD proposed by Evans~\cite{Evans:1982aa} without considering quantum nuclear effects. More details can be found in Ref.~\onlinecite{csld}.} and  experimental data reported by Kosaka~\etal\cite{Kosaka2017PCCP}, Suekuni~\etal\cite{Suekuni2018} and Chetty~\etal\cite{Chetty2015PCCP}, although discrepancies are found in the magnitude of $\kappa_{L}$. It should be noted that large discrepancies exist among experimental $\kappa_{L}$, which can be probably ascribed to the well-known experimental challenges in (i) precisely ensuring stoichiometry~\cite{Skinner1972,Makovicky:1978aa}, (ii) accurately determining the electronic contribution of the thermal conductivity \footnote{Cubic tetrahedrites have been both theoretically and experimentally confirmed to be metallic (two holes per formula unit)~\cite{Lu2013}. Therefore experimentally measured thermal conductivity contains large electronic contribution, which is usually estimated using the Wiedemann-Franz law and subtracted from the measured total thermal conductivity to obtain the lattice contribution.} and (iii) extreme sensitivity of $\kappa_{L}$ to Cu off-stoichiometry in \CuSbS~\cite{Vaqueiro:2017aa}. In this regard, our calculation establishes an independent theoretical estimation, which could be further examined by carefully designed experiments. Meanwhile, our theoretical $\kappa_{L}$ might be enhanced by including anharmonic contributions to the heat-flux operator~\cite{Hardy1963,Sun2010}, which are naturally included in MD simulations but absent in our model.

Finally, we conjecture that the quartic anharmonicity might also play a key role in the unusual electrical transport properties of \CuSbS. Experiments have provided evidence for a variable-range hopping mechanism of electronic conductivity at high $T$ as well as for a low-$T$ metal-insulator transition that persists even in stoichiometric crystals~\cite{Suekuni_2012,Lu2013}. Investigation of the structural disorder and electron-phonon interactions associated with the rattling-like Cu(2) vibrations, and possibility of off-diagonal tunneling contributions to the electrical current should provide a deeper understanding of these phenomena.


\begin{acknowledgments}
\textbf{Acknowledgments: } Y.X. and C.W. acknowledge financial support received from (i) Toyota Research Institute (TRI) through the Accelerated Materials Design and Discovery program (thermal conductivity calculations), (ii) the Department of Energy, Office of Science, Basic Energy Sciences under grant DE-SC0014520 (theory of anharmonic phonons), and (iii) the U.S. Department of Commerce and National Institute of Standards and Technology as part of the Center for Hierarchical Materials Design (CHiMaD) under award no. 70NANB14H012 (DFT calculations). V.O. acknowledges financial support from the National Science Foundation Grant DMR-1611507. This research used resources of the National Energy Research Scientific Computing Center, a DOE Office of Science User Facility supported by the Office of Science of the U.S. Department of Energy under Contract No. DE-AC02-05CH11231.
\end{acknowledgments}
\bibliography{CuSbS}

\end{document}